\documentstyle[prl,aps,graphics,floats,preprint]{revtex}
\begin{document}

\draft

\title{Current induced spin wave excitations in a single ferromagnetic 
layer}
\author{Y.~Ji and C. L.~Chien}
\address{Department of Physics and Astronomy, The Johns Hopkins University, 
Baltimore, Maryland 21218}
\date{\today}
\maketitle

\begin{abstract}
A new current induced spin-torque transfer effect has been observed in a 
$single$ ferromagnetic layer without resorting to multilayers. At a specific 
current density of one polarity injected from a point contact, abrupt 
resistance changes due to current-induced spin wave excitations have been 
observed. The critical current at the onset of spin-wave excitations 
depends linearly on the external field applied perpendicular to the 
layer.  The observed effect is due to current-driven heterogeneity in an 
otherwise uniform ferromagnetic layer.
\end{abstract}


%--------------------------------------------------------------------------

Recently, spin-torque transfer effects have attracted a great deal of 
attention due to potential device applications~\cite{Slon97} as 
well as the novel fundamental physics it reveals. The magnetic 
configuration of a system is known to affect its electrical behavior, 
such as those in anisotropic, giant~\cite{Baibich}, and 
tunneling~\cite{Moodera} magnetoresistance (AMR, GMR and TMR 
respectively) effects. Spin-torque transfer is an example of the reverse effect, 
where an electrical current can alter the magnetic configuration of a 
system. It has been theoretically predicted~\cite{Berger,Slon96,Bazaliy} 
and experimentally observed~\cite{Tsoi98,Tsoi00,Myers,Katine} that at 
sufficiently high current densities, a spin-polarized current is able to 
exert a torque on a ferromagnetic entity and switch its magnetization in 
the low field regime or stimulate spin precession (spin waves) in the 
same system in the high field regime. Tsoi $et~al.$~\cite{Tsoi98,Tsoi00} 
first reported on spin wave excitations in Co/Cu multilayers by current 
injection through a mechanical point contact. Myers $et~al.$~\cite{Myers} 
observed current induced switching in point contacts made by 
nano-lithography on Co/Cu/Co trilayers. Both magnetization switching and 
spin precession have been reported by Katine $et~al.$~\cite{Katine} in 
patterned Co/Cu/Co trilayered nanopillars. Other current induced effects 
have also been observed in Ni nanowires~\cite{Wegrowe} and manganite 
junctions~\cite{Sun}.

To date, most theoretical treatments and experimental observations have 
featured heterostrctures such as FM/NM multilayers or FM/NM/FM trilayers, 
where FM=ferromagnet and NM=non-magnetic metal, with a current flowing 
perpendicular to the layers. There are two considerations that favor the 
multilayer structures. First of all, these structures provide at least one 
static FM layer and one free FM layer in the system. The static layer has 
a fixed magnetization throughout the measurement (realized by shape 
anisotropy and/or the application of an external field).  The static 
layer, acting as a polarizer, also defines the polarization direction for 
the current. The free layer is separated from the static layer by a 
non-magnetic metal. As the spin-polarized current produced by the static 
layer passes through the free layer, the current carriers transfer their 
spin angular momenta onto the magnetic moments in the free layer, thereby 
imparting a torque. The consequence of the torque depends on the polarity 
of the current. If the electrons flow from the static layer to the free 
layer, the torque favors a parallel alignment of the FM layers.  For 
electrons flowing in the opposite direction, the torque favors an 
anti-parallel alignment of the FM layers.  As the current is swept between 
polarities, at a critical current value, the magnetization of the free 
layer can be altered between parallel and anti-parallel alignment with 
respect to the static layer. In the presence of a strong external field 
(of the order of tesla), however, full magnetization reversal into 
anti-parallel state becomes unfeasible. Instead, spin waves are 
stimulated, featuring a precession of spin moments between the parallel 
and the anti-parallel states.

Secondly, the FM/NM/FM trilayer or FM/NM multilayer structure also 
exhibits GMR, which becomes a detection mechanism for the magnetization 
reversal or spin precession. The parallel and the antiparallel 
configurations of the FM layers define the low and high resistance states 
respectively in the system. Deviations in configuration, such as that of 
magnetization reversal, result in changes in the resistance. The spin wave 
precession can be identified as an increase in the resistance ($V$/$I$) or a 
peak in differential resistance at a certain critical current value in the 
polarity such that the electrons flow from the free layer to the static 
layer.

The persuasive rationale for a multilayer structure notwithstanding,  
multilayer structures have been exclusively employed in experiments for 
realizing the spin-torque effects. We report in this Letter experimental 
observation of current induced spin waves in a $single$ ferromagnetic layer 
without heterostructures involving non-magnetic metals. This effect is due 
to the rapidly degrading current density, which results in magnetic 
inhomogeneity with a large magnetization gradient within the single layer.

Mechanical point contact technique~\cite{Jasen} has been used to inject 
a current with a current density in excess of 10$^9$ A/cm$^2$ from a silver 
tip into a 3000~\AA~sputtered Co layer, with an external field 
up to 9~T applied perpendicular to the Co layer. All the measurements we report 
in this Letter have been carried out at 4.2~K, using a four-probe 
method, with $I$+ and $V$+ made on the tip and $I$- and $V$- made on the 
film. For current in the negative polarity, electrons are flowing from the tip 
into the Co film. We have measured resistance $R$ and differential 
resistance $dV/dI$ either as a function of current $I$ at a fixed field $H$ 
or as a function of field $H$ at a fixed bias current $I$. The differential 
resistance $dV/dI$ has been measured using a phase sensitive detection 
technique with an AC modulation current of 1~$\mu$A and 2~kHz.

Figure 1 shows a typical contact that exhibits the unusual behavior 
of $R$ and $dV/dI$. The contact resistance is about 28~$\Omega$ at zero bias 
and the external field is 5~T. In the positive polarity, both quantities are 
slowly varying as a function of current $I$. The small increase of $R$ or 
$dV/dI$ at higher bias can be attributed to phonon and magnon 
scattering~\cite{Jasen}.  At negative polarity, however, a sharp peak in 
$dV/dI$ and a prominent step in $R$ (enlarged in the inset) can be seen at 
a bias current of $I$~=~-3.27 mA. The upward jump of $R$ is about 1~$\Omega$, 
about 3\% of the total resistance. The differential resistance changes by 
more than 100\% at the same bias. Beyond the main peak, $dV/dI$ also 
displays a small upward step at even higher bias. We have studied more than 
50 point-contacts that show similar features. The peak in $dV/dI$ or step 
in $R$ is always present in the negative bias, and never in the positive 
bias. The peak structure disappears at a field lower than 4$\pi$$M$ 
(1.7~T), the demagnetizing field of the Co film.

Figure 2(a) shows the results of an Ag/Co contact at different external 
fields from 2~T to 9~T.  All the curves display a peak structure but at 
different negative bias current. All the curves fall onto the same 
background, demonstrating that the contact is not altered throughout the 
measurements. We define the current value corresponding to the peak in 
$dV/dI$ as the critical current $I_c$, which, as shown in Figure 2(b), 
depends linearly on $H$.

It is interesting to note that Figure 2(b) establishes also a phase diagram 
for that contact, excluding the low field region where spin waves could 
not be stimulated. The region above the line represents the excited states 
with spin precession whereas the region below the line is the ground state 
where all the spin moments are aligned by the external field. The 
measurements performed in Figure 2(a) correspond to a scan along a vertical 
line at a fixed $H$ in Figure 2(b). As the boundary defined by the straight 
line is crossed, resistance changes (a step in $R$ and a peak in $dV/dI$) are 
encountered. The phase diagram so determined is specific to a given contact.

The phase diagram in Figure 2(b) indicates that one can also perform a field 
scan along a horizontal line and monitor resistance and $dV/dI$ as the phase 
boundary is crossed.  The result of such a scan using a different contact 
is shown in Figure 3.  At a field of 5~T, as the current is scanned from 
-5~mA to -3~mA, as shown in Figure 3(a), a peak appears in $dV/dI$ at -3.8~mA, 
signifying the excitation of spin waves beyond this critical value.  We 
then hold the current at -3.0 mA, ramp down the field, and measure $R$ and 
$dV/dI$ (Figure 3(b-c)). At 2.6~T, a peak in $dV/dI$ and a step in $R$ appear. 
As shown in Figure 2, as the field is decreased, the critical current 
value will also decrease. At a lower field, 2.67~T in this case, the 
critical current value becomes -3.0 mA, hence the $dV/dI$ peak.

We note that the results shown in Figure 1-3 with a single Co layer bear 
close resemblance to the experimental results of Tsoi 
$et~al.$~\cite{Tsoi98,Tsoi00} and Katine $et~al.$~\cite{Katine} using 
multilayers. The characteristics also agree well with the theory of 
Slonczewski~\cite{Slon99} in the context of multilayers. The single Co layer 
geometry that we have used does not contain a static layer/NM/free layer 
geometry nor a GMR detection mechanism.  The multilayer geometry is therefore 
not a prerequisite for observing the spin-torque transfer effect.  In the 
following, we propose a microscopic picture that accounts for the observed 
effects, and the salient differences from those observed in multilayers.

The geometry that we use for spin injection through a point contact is 
schematically in Figure 4.  In the negative polarity, where the spin waves 
are excited, electrons are flowing from the tip into the Co film. The 
contact size can be estimated from the contact resistance $R$, the resistivity 
$\rho$, and the electron mean free path $l$ using 
$R$=4$\rho$$l$/3$\pi$$a^2$+$\rho$/2$a$, 
the Wexler formula,~\cite{Jasen,Wexler} which is a combination of Sharvin 
and Maxwell resistances~\cite{Wexler}.  The typical resistance value of 
our contacts is between 10-100 $\Omega$, corresponding to a contact radius 
of the order of 3-8 nm.  As an estimate, at an injection current of 3 mA 
through such a small contact, the current density is between 
1.2$\times$10$^9$ A/cm$^2$ and 9$\times$10$^9$ A/cm$^2$. However, immediately 
after the current injection into the Co film, the current spreads and the 
current density decrease rapidly due to the expanding geometry. Beyond a 
certain depth, which is marked by a horizontal dashed line in Figure 4, 
the current density would be too low to excite spin waves. All the spin 
moments below would be unaffected by the current and aligned in the 
external field direction. The region below serves as a static region in 
the system and provides a source for the spin-polarized carriers. The 
region above the boundary, where the current density is sufficiently high, 
acts as the free region. In this manner the single Co layer is separated 
into a free "layer" immediately below the point contact and a fixed layer 
for the remainder of the Co layer. The thickness of the "free" layer is 
determined by the current distribution underneath the point contact. In 
this sense, the single Co film is divided into two regions by a dynamic 
process, rather than physically separated by the insertion of a 
non-magnetic layer as in Co/Cu/Co trilayers.

Electrons from the tip first flow through the free region and then enter 
the static region. As discussed earlier, in this polarity the current 
induced torque onto the free region tends to deflect the spin moments from 
the parallel alignment, causing spin precession. The resistance step or 
$dV/dI$ peak, which is considered the signature of spin precession, can be 
understood as such. As the precession is excited, spin moments in the free 
region will have a small tilt angle with respect to the spin in the static 
region.  Given the small length scale, the magnetization gradient would be 
extremely high. As the charge carriers moving from the free region to the 
static region, their spin moments could not relax adiabatically. Instead, 
spin-dependent scattering occurs, giving rise to a high resistance state.

The above model accounts for the experimental results using a single FM 
layer well. It also implies certain features that are different from 
multilayers. Slonczewski~\cite{Slon99} has calculated the current threshold 
of spin excitations in the context of point contact current injection into a 
laterally unbounded FM1/NM/FM2 structure, with a field perpendicular to 
plane. There are two distinct origins of current threshold requirements. 
One is due to energy loss through radiation from the central excited 
region under the contact to the rest of the FM1 layer. This term is 
proportional to exchange stiffness and the thickness of FM1 layer. 
The other is attributed to the alignment of the effective field, which 
includes the external field, the demagnetizing field, the anisotropy 
field and the exchange field from FM2 layer. The current induced torque has 
to be large enough to overcome the effective field and viscous damping. So 
the field dependent term is proportional to the effective field, the Gilbert 
damping coefficient, and the total magnetic moments in the excitation region.
In the case of a single Co layer, both terms imply a higher threshold current 
than those in multilayers. The propagation of energy will no longer be 
constricted in a very thin FM1 layer. Instead the excited region will 
dissipate energy through exchange interaction into the entire 300 nm thick 
Co film. The exchange field experienced by the excited region is also much 
larger, without separation by a non-magnetic metal between the free layer 
and the static layer.

The $dV/dI$ peaks that we have observed occur in a negative bias voltage 
between 100~mV to 200~mV in Figure 2(a). The estimated current density is 
5-8$\times$10$^9$ A/cm$^2$, which is considerably higher than the value of 
10$^8$ A/cm$^2$ in trilayered pillars~\cite{Katine}, and the value of about 
1$\times$10$^9$ A/cm$^2$ for multilayers~\cite{Tsoi98,Tsoi00}. In the latter, 
since the current injected from a point contact spreads out in the multilayers, 
perhaps only the top layers can be excited and the rest of the layers remain 
fixed as a polarizer.

In conclusion, we report spin wave excitations in a single ferromagnetic 
layer by high-density current injection through a point contact. Without a 
multilayer structure involving non-magnetic metal or a 
nanolithographically patterned trilayer nano-pillar, we have observed spin 
precession, albeit at a higher current density. The spin precession occurs 
in a constricted region underneath the point contact. The resistance 
changes, signatures of spin-wave excitations, are attributed to spin 
dependent scattering due to a localized large magnetization gradient.  The 
effects that have been observed in single ferromagnetic layers imply that 
similar effects could also appear in multilayers, even though the 
multialyers have been designed to create magnetic heterogeneity.

Work supported by National Science Foundation grant no. DMR00-80031 and 
DMR01-01814.

%----------------------------------------------------------------------------

\begin{figure}[p]
\centering
\resizebox*{5in}{!}{\includegraphics*{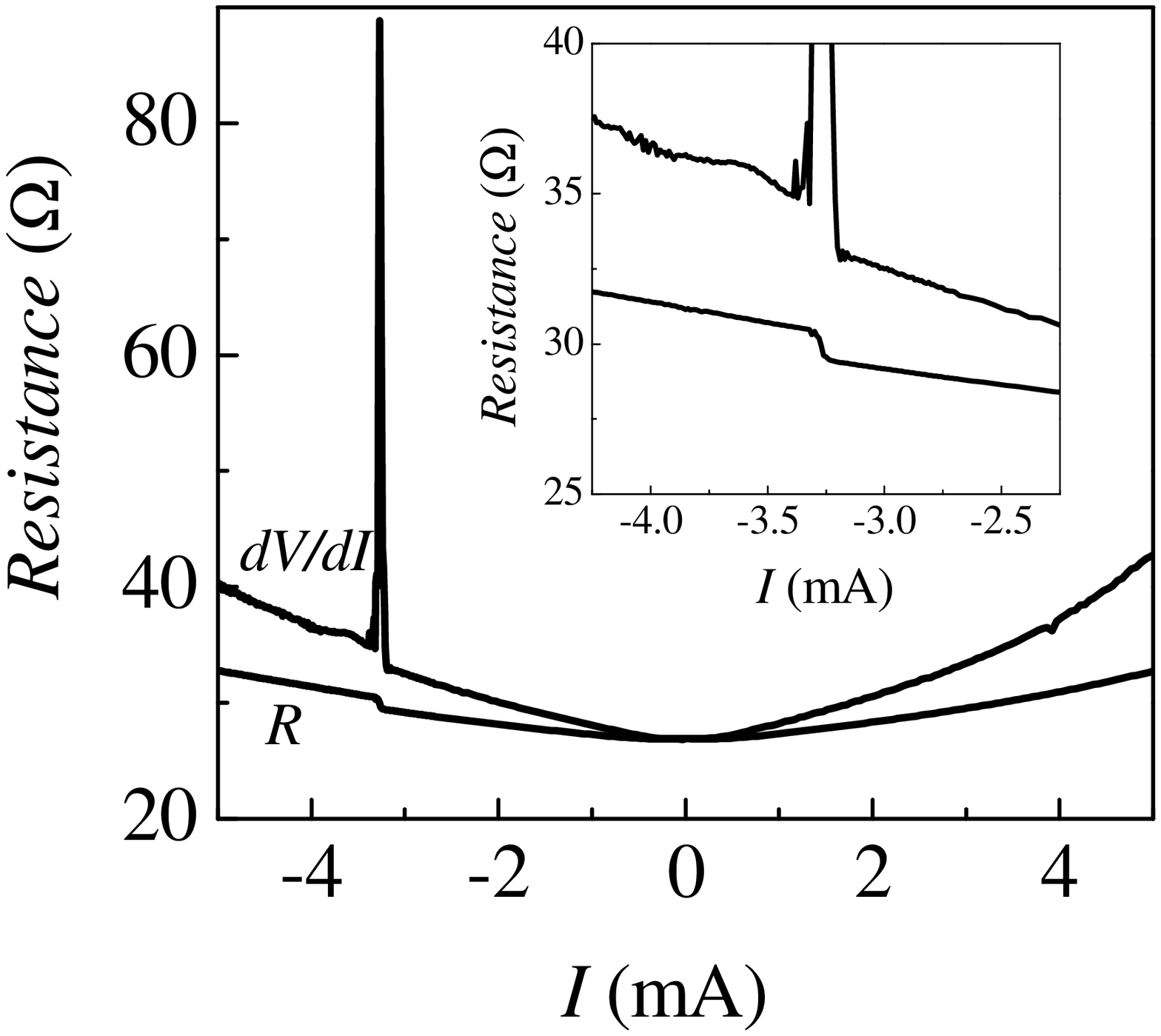}}
\caption{The $R$-$I$ and $dV/dI$-$I$ plots for a point contact between an 
Ag tip and a Co film subject to an external field of 5~T perpendicular to 
the Co layer at 4.2~K, showing the peak (or step) at bias current 
$I$=- 3.27~mA, enlarged in the inset.}
\label{Fig1}
\end{figure}

\begin{figure}[p]
\centering
\resizebox*{5in}{!}{\includegraphics*{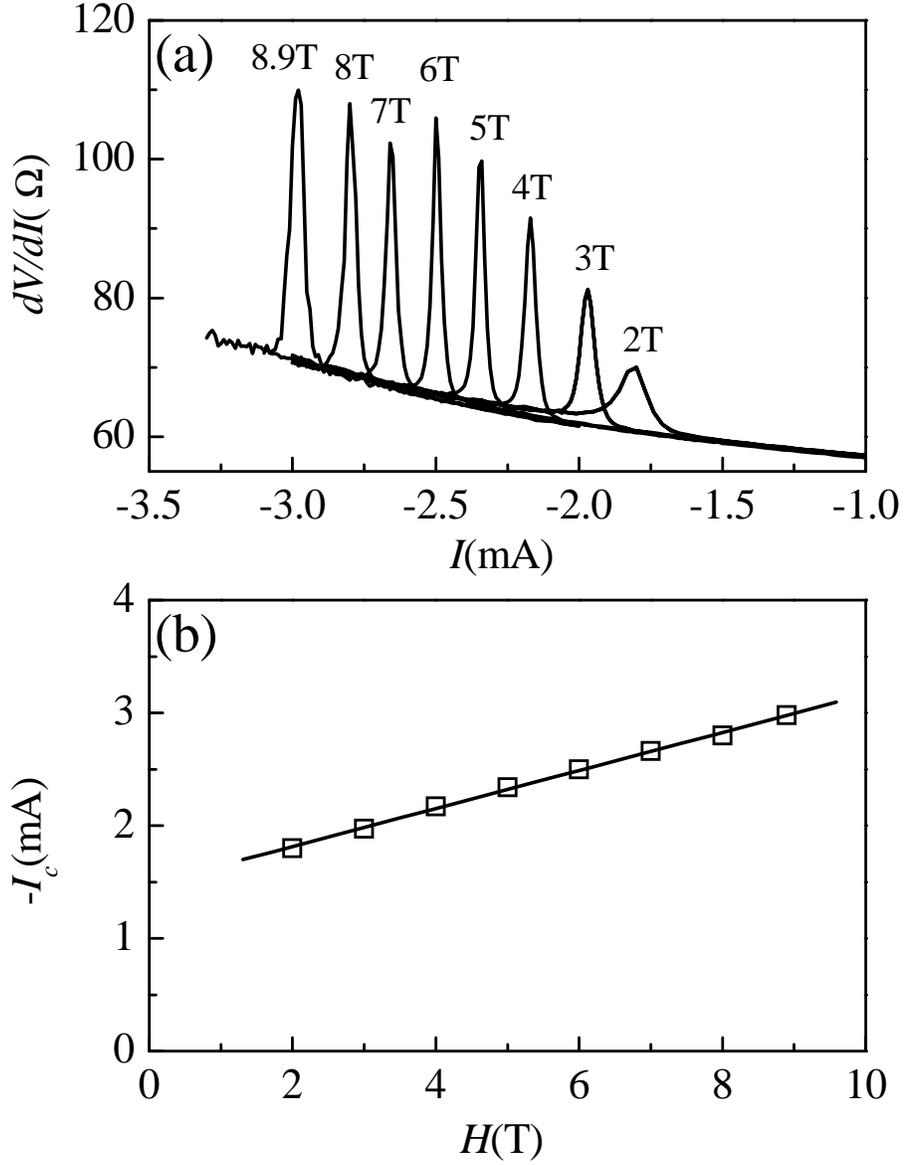}}
\caption{(a) The $dV/dI$-$I$ plots at different fields for an Ag/Co point 
contact at 4.2~K. The current value at the peak position, defined as $I_c$, 
depends linearly on the external field as shown in (b).}
\label{Fig2}
\end{figure}

\begin{figure}[p]
\centering
\resizebox*{5in}{!}{\includegraphics*{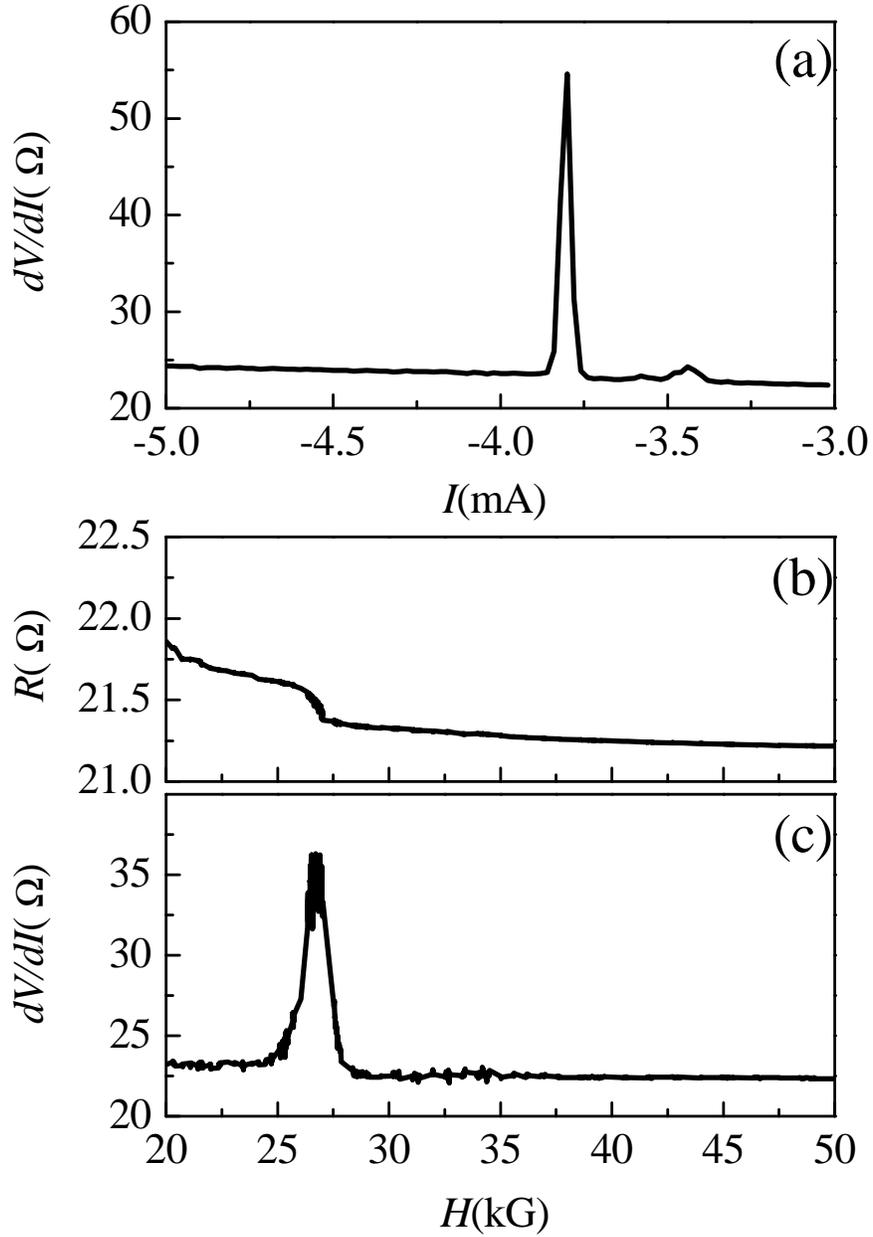}}
\caption{Resistance and $dV/dI$ of an Ag/Co point contact at 4.2 K. (a) 
Vary the current from - 5 mA to - 3mA in a fixed field of 5~T, (b)(c) Vary 
the magnetic field at a fixed current of - 3 mA.}
\label{Fig3}
\end{figure}

\begin{figure}[p]
\centering
\resizebox*{5in}{!}{\includegraphics*{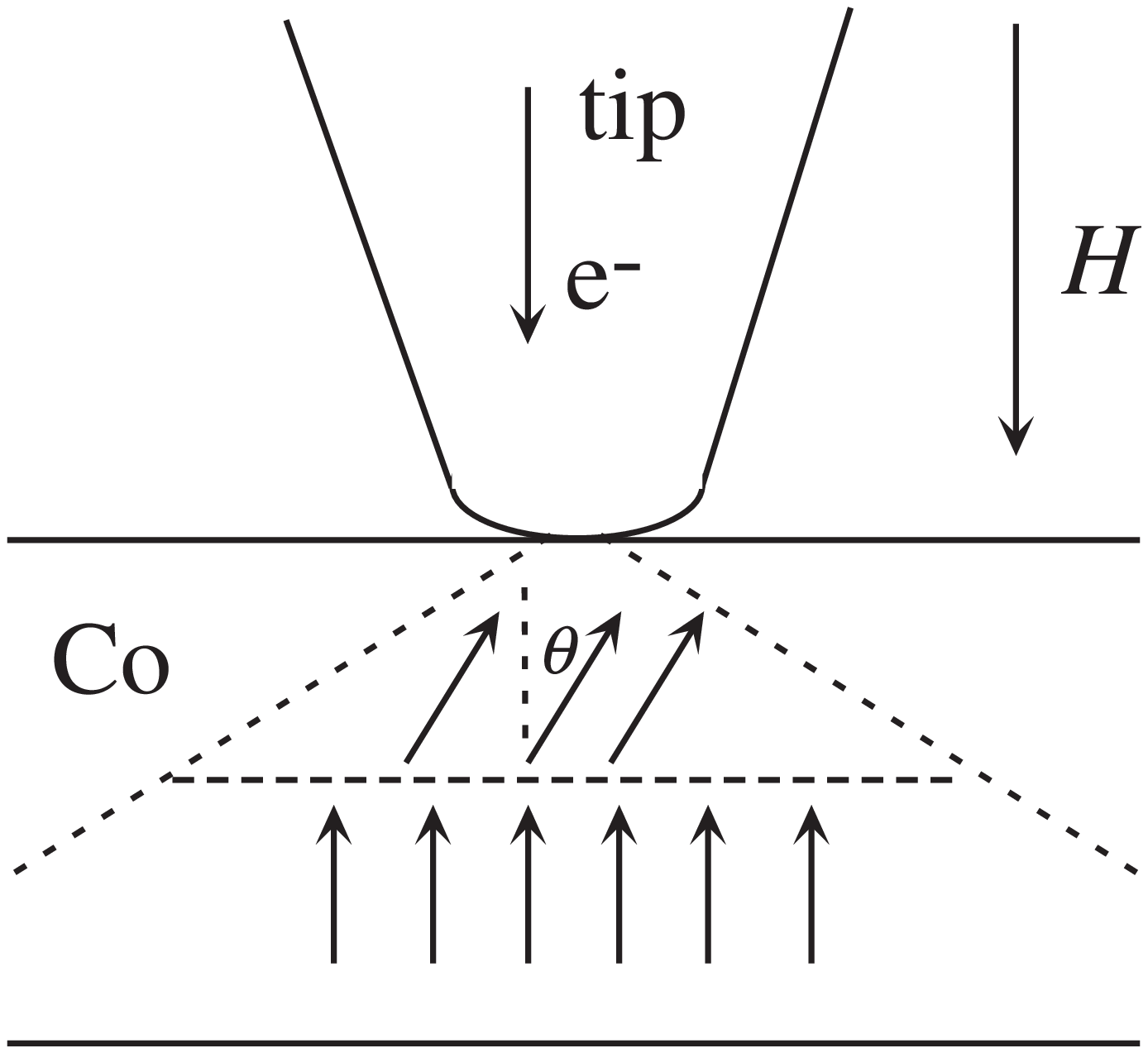}}
\caption{A microscopic picture of a point contact between an Ag tip and a 
Co film with an external magnetic field applied perpendicular to the Co 
layer. On entering the Co film, electrons first pass through a localized 
"free region" right underneath the tip and before entering the "static 
region" as the current spread out. The horizontal dashed line mark the 
boundary between the free and static regions.}
\label{Fig4}
\end{figure}

\end{document}